\documentclass[
 reprint,
 nofootinbib,
 showkeys,
 amsmath,
 amssymb,
 aps,
 prb,
 floatfix,
 superscriptaddress
 ]{revtex4-2}

\usepackage{graphicx}
\usepackage{multirow}
\usepackage{xcolor}
\usepackage{dcolumn}
\usepackage{bm}
\usepackage{physics}
\usepackage{hyperref}

\newcommand{\Hc}[0]{\text{H.c.}}

\begin{document}

\title{Signatures of the order parameter of a superconducting adatom layer in magnetic field-dependent quasiparticle interference}

\author{B.~A.~Levitan}
\email{benjamin.levitan@weizmann.ac.il}
\affiliation{Department of Condensed Matter Physics, Weizmann Institute of Science, Rehovot 7610001, Israel}
\affiliation{Department of Physics and the Centre for the Physics of Materials, McGill University, 3600 rue University, Montr\'{e}al, Qu\'{e}bec, H3A 2T8, Canada}
\author{J.~Eid}
\author{T.~Pereg-Barnea}
\affiliation{Department of Physics and the Centre for the Physics of Materials, McGill University, 3600 rue University, Montr\'{e}al, Qu\'{e}bec, H3A 2T8, Canada}

\date{\today}

\begin{abstract}
Experiments have observed superconductivity in atomically-thin metallic layers deposited on semiconducting substrates. As in any superconductor, it is important to determine the structure of the superconducting pairing function in order to reveal the mechanism responsible for superconductivity. To that end, we study the possible superconducting states of two-dimensional triangular lattices. We calculate the quasiparticle interference (QPI) patterns which would result from various nearest-neighbor pairing order parameters, and show how the QPI can be used to distinguish between those order parameters. The QPI patterns are the momentum-space representations of real-space local density-of-states fluctuations: the QPI signal at momentum $q$ reveals the strength of scattering processes at that momentum transfer. We show how characteristic differences between scattering from charge disorder (i.e.~impurities) and from order-parameter disorder (i.e.~vortices) can be used to identify the angular momentum of the superconducting pairs.
\end{abstract}

\maketitle

\section{Introduction}
Recent advances in the preparation of ultra-clean surfaces have made new two-dimensional materials possible, which are difficult or impossible to create in free-standing form. In particular, by depositing a controlled amount of a chosen adatom onto an otherwise-pristine surface, and annealing from high temperatures to facilitate lattice relaxation, high-quality superlattices can be formed.

An interesting example of such a system is obtained by depositing tin atoms onto a (111) surface of silicon, in a ratio of one tin atom per three silicon unit cells. The resulting surface hosts a single partially-filled electronic band, which exists inside the silicon bandgap. At half-filling (one unpaired electron per tin site), the surface has been observed to show Mott insulating behavior \cite{Li2013:magnetic_surface_order}. When mobile holes are introduced into the Mott insulator (by boron-doping the silicon substrate), superconductivity has been observed \cite{Ming3,Wu2020:holedoped_adatom}. This behavior resembles that of the high-$T_c$ superconductors, whose enigmatic superconducting states also emerge upon doping Mott insulators \cite{Lee2006}. The proximity of the superconductor to a magnetic insulating phase suggests that the superconductivity in the adatom system could likewise be unconventional.

Unfortunately, when nature presents us with a superconductor, it does not come with a simple label indicating the underlying pairing mechanism. Therefore, whenever a new superconductor is discovered, the question of whether it is conventional (BCS-like, mediated by phonons \cite{BCS}) or unconventional (mediated by electronic correlations \cite{Raghu}) arises immediately. It is commonly believed that electron-phonon interactions result in $s$-wave superconductivity, while electron-electron (Coulomb) interactions yield pairing with higher orbital angular momentum ($\ell$). Determining the symmetry of the order parameter is therefore a critical step in identifying the pairing mechanism. As well, order parameters with even $\ell$ correspond to spin-singlet Cooper pairs, while those with odd $\ell$ correspond to spin-triplet pairs. Because spin-triplet pairing is a necessary (but not sufficient) condition for topological superconductivity when spin-orbit coupling is absent or weak, identifying the order parameter symmetry is also necessary for applications which rest on topological superconductivity, such as quantum information processing based on Majorana modes \cite{Kitaev2003:anyons, Oreg2010:Majoranas, Lutchyn2010:Majoranas}.

In this work, we propose an experiment to reveal the order parameter symmetry in the tin-on-silicon adatom system, and related materials, through quasiparticle interference (QPI) patterns \cite{Hoffman}. One obtains these patterns as the Fourier transform of local density-of-states (LDOS) modulations which result from scattering due to disorder, measured through scanning-tunneling spectroscopy (STS). We consider two types of scatterer in an otherwise-clean superconductor: a charge impurity, and a vortex. A charge impurity is modeled as a point-like perturbation in the chemical potential, while a vortex is modeled as a local suppression of the pairing function.  We consider various superconducting pairing functions which are defined on nearest-neighbor links of our hexagonal lattice model, and demonstrate that the QPI resulting from the two types of perturbations can help distinguish between the different possible gap functions. In particular, we highlight how depending on the order parameter symmetry, certain transitions are suppressed in the vortex scattering QPI --- those involving a sign change in the superconducting gap function --- and the signals corresponding to those transitions will therefore not increase with magnetic field. Therefore, the magnetic field dependence of the QPI is sensitive to the phase winding of the order parameter.

The paper is organized as follows. We introduce a tight-binding model for the superconducting tin-on-silicon system in Sec.~\ref{sec:model}, describing the clean limit in Sec.~\ref{subsec:clean_limit}, and the scatterers in Sec.~\ref{subsec:scatterers}. We then compute the QPI patterns in Sec.~\ref{sec:qpi}. We analyze the on-shell scattering contributions to the QPI in Sec.~\ref{subsec:qpi_onshell}, and show how they can be used to distinguish between different order parameters in Sec.~\ref{subsec:identifying_delta}. We then summarize our findings and conclude in Sec.~\ref{sec:conclusion}. 

\section{The model} \label{sec:model}
\subsection{The clean limit} \label{subsec:clean_limit}
\begin{figure}
    \centering
    \includegraphics[width=0.8\columnwidth]{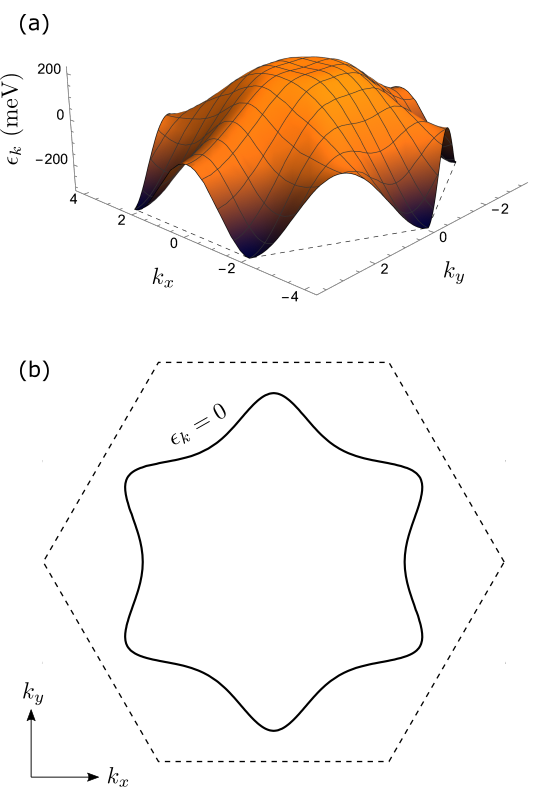}
    \caption{\label{fig:dispersion} (a) The bare dispersion $\epsilon_k$, and (b) its warped hexagonal Fermi surface, within the first hexagonal Brillouin zone (dotted lines).}
\end{figure}

We describe the normal state of the metallic atomic monolayer using a tight-binding model on the triangular lattice, including several beyond-nearest-neighbour hopping terms, and neglecting magnetic ordering. The corresponding momentum dispersion $\epsilon_k$ is given explicitly in Appendix \ref{appendix:dispersion+order_params}, Eq.~\eqref{eq:epsilon}. For the parameters relevant to atomic Sn deposited on a Si (111) surface, obtained \textit{ab-initio} by the authors of Ref.~\cite{Li2013:magnetic_surface_order}, $\epsilon$ has a warped hexagonal Fermi surface near charge neutrality, shown in Fig.~\ref{fig:dispersion}(b).

Our goal is to point out an experimental probe which could identify the superconducting order parameter. Therefore, we make no attempt to explicitly include the interactions which give rise to superconductivity --- other authors have taken up that task \cite{Wolf2022:triplet_superconductivity} --- and instead we simply write down an effective single-particle Hamiltonian including a static (non-self-consistent) order parameter $\Delta_{k}$. We will then work out the consequences of several choices of $\Delta_{k}$ with respect to STS experiments. 

Using the standard Bogoliubov-de Gennes (BdG) formalism \cite{Bogoljubov}, in the clean limit, the Hamiltonian is
\begin{equation}
    H_0 = \sum_{k} \Psi^{\dag}_{k} h_{k} \Psi_{k}.
\end{equation}
$\Psi_k$ is a two-component pseudospinor in particle-hole space,
\begin{equation}    \label{eq:singlet_spinor}
    \Psi_k
    =
    \begin{pmatrix}
        c_{k, \uparrow} \\
        c^{\dag}_{-k, \downarrow}
    \end{pmatrix}
\end{equation}
when considering spin-singlet order parameters, and
\begin{equation}    \label{eq:triplet_spinor}
    \Psi_k
    =
    \begin{pmatrix}
        c_{k, \uparrow} \\
        c^{\dag}_{-k, \uparrow}
    \end{pmatrix}
\end{equation}
for spin-triplet order parameters\footnote{The most general triplet pairing is $\sum_{k \alpha \beta} c^{\dag}_{k \alpha} \left( \Vec{\Delta}_k \cdot \Vec{S} \right)_{\alpha \beta} c^{\dag}_{-k \beta} + \Hc$. Taking $\Vec{\Delta}_k$ to point along the spin quantization axis yields $\Delta_k S_z$, meaning that spin-$\uparrow$ and spin-$\downarrow$ electrons experience order parameters with equal magnitude and opposite sign. This  sign is not observable in spin-insensitive measurements, so we can safely analyze only one component, which we take to be spin-$\uparrow$. Absent spin-orbit coupling, this choice can be made without loss of generality.}. $c_{k \sigma}$ annihilates an electron with momentum $k$ and spin $\sigma$. $h_k$ is the $2 \times 2$ BdG Hamiltonian,
\begin{equation}    \label{eq:BdG_hamiltonian}
    h_k
    =
    \begin{pmatrix}
        \epsilon_k  &   \Delta_k    \\
        \Delta^*_k  &   - \epsilon_k
    \end{pmatrix}.
\end{equation}
Diagonalizing $h_k$ yields the quasiparticle bands, with energies
\begin{equation}
    \pm E_k = \pm \sqrt{\epsilon_k^2 + | \Delta_k |^2}.
\end{equation}
Note that throughout this paper, we work in units where $\hbar = \text{lattice spacing} = 1$.

\begin{figure}[t]
    \centering
    \includegraphics[width=0.6\columnwidth]
        {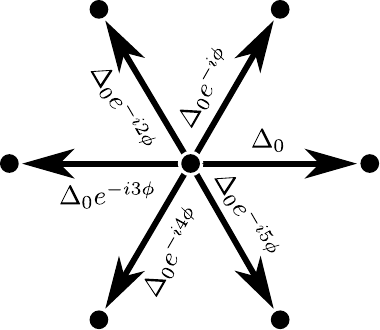}
    \caption{For pairing with angular momentum $\ell$, the pairing phase winds by $2 \pi \ell$ under a full rotation, corresponding to $\phi = 2 \pi \ell / 6$.}
    \label{fig:pairing_phases}
\end{figure}

For the metal-on-semiconductor layers under consideration, the superconducting state is observed upon doping a Mott insulator \cite{Wu2020:holedoped_adatom}. It was suggested that the Mott physics may prevent on-site $s$-wave pairing, and give rise to unconventional superconductivity \cite{Ming2,Wu2020:holedoped_adatom, Wolf2022:triplet_superconductivity}. We therefore consider the first four angular momentum channels of nearest-neighbour pairing, labelled by $\ell = 0, \dots, 3$ (respectively extended $s$-, chiral $p$-, chiral $d$-, and $f$-wave). Recent theory suggests that each of $\ell = 1, 2, 3$ may occur in some region of the phase diagram as a function of doping and nearest-neighbour repulsion \cite{Wolf2022:triplet_superconductivity}. Because fermions anticommute ($c_1 c_2 = - c_2 c_1$), pairing with even $\ell$ (odd $\ell$) must occur in the spin-singlet (spin-triplet) channel. $\ell$ labels how many times the phase of the order parameter winds under a $2 \pi$ rotation. Since each site of the triangular lattice has six nearest-neighbour bonds (separated by an angle of $2 \pi / 6$), the pairing terms on bonds related by a $2 \pi / 6$ rotation have a relative phase of $2 \pi \ell / 6$, as depicted in Fig.~\ref{fig:pairing_phases}. In that figure, arrows point from a site $r$ to its neighbours $r + \delta$, indicating pairing terms $\Delta_{\delta} c^{\dag}_{r + \delta, \uparrow} c^{\dag}_{r, \downarrow} + \Delta_{\delta}^* c_{r, \downarrow} c_{r + \delta, \uparrow}$ in the spin-singlet cases (even $\ell$, $\Delta_{\delta} = \Delta_{-\delta}$), and $\Delta_{\delta} c^{\dag}_{r + \delta, \uparrow} c^{\dag}_{r, \uparrow} + \Delta_{\delta}^* c_{r, \uparrow} c_{r + \delta, \uparrow}$ in the spin-triplet cases (odd $\ell$, $\Delta_{\delta} = - \Delta_{-\delta}$). The explicit form of each order parameter in momentum space is given in Appendix \ref{appendix:dispersion+order_params}, Eqns.~\eqref{eq:extended_swave} through \eqref{eq:fwave}. In general, the order parameter can take a more complicated form (e.g.~involving pairing between next-nearest-neighbors). However, pairing between nearest-neighbors is likely dominant, and our analysis should remain valid with the addition of further pairing terms of the same angular momentum. 

\subsection{Scatterers} \label{subsec:scatterers}
Moving away from the clean limit, we consider two kinds of elastic scatterer. The first is static charge disorder, corresponding to a potential $u_c (r)$:
\begin{equation}    \label{eq:charge_scatterer}
    H_{c}
    =
    \sum_{r} \sum_{\sigma} 
    u_c (r) c^{\dag}_{r \sigma} c_{r \sigma}
    =
    \frac{1}{\sqrt{N}}
    \sum_{k q} u_{c,q} \Psi^{\dag}_{k} \sigma_3 \Psi_{k - q},
\end{equation}
where $\sigma_3 = \begin{pmatrix} 1 & 0 \\ 0 & -1 \end{pmatrix}$ is the $3^{rd}$ Pauli matrix in particle-hole space, and $u_{c,q} = \frac{1}{\sqrt{N}} \sum_{r} e^{-i q \cdot r} u_c (r)$ is the Fourier transform of $u_c (r)$ on a periodic lattice with $N$ sites.

The second type of scatterer we consider is a static arrangement of superconducting vortices. We treat a vortex in the simplest possible way, which neglects its topological character (i.e.~phase winding), retaining only the fact that superconductivity is suppressed in the vortex core. To that end, we model a vortex as a local perturbation where the magnitude of pairing is modulated on all links $\delta$ connecting to a single site \cite{PeregBarnea2008:QPI, Plamadeala2010:vortices}. This approach has proven useful in interpreting the results of QPI experiments under magnetic field in the cuprates \cite{HanaguriCuprateQPI} and the iron pnictides \cite{HanaguriIronQPI}. Many vortices together are then built out of such local perturbations weighted by a function $u_v (r)$ with Fourier transform $u_{v,q}$:
\begin{multline}    \label{eq:vortex_scatterer}
    H_{v}
    =
    \sum_{r} \sum_{\delta}
    u_v (r) \chi (\delta) \left( 
    c^{\dag}_{r + \delta, \uparrow} c^{\dag}_{r, \downarrow}
    + c^{\dag}_{r, \uparrow} c^{\dag}_{r - \delta, \downarrow}
    \right) + \Hc   \\
    =
    \frac{1}{\sqrt{N}}
    \sum_{k q} u_{v,q} \Psi^{\dag}_{k} V^{v}_{k, k-q} \Psi_{k - q}
\end{multline}
in terms of the matrix
\begin{equation}    \label{eq:vortex_matrix}
    V^{v}_{k, k-q}
    =
    \begin{pmatrix}
        0                               &   \chi_{k} + \chi_{k-q}   \\
        \chi^{*}_{k} + \chi^{*}_{k-q}   &   0
    \end{pmatrix}.
\end{equation}
$\chi_k$ is a dimensionless function giving the momentum dependence (or link dependence in real space) of the order parameter, $\Delta_k = \Delta_0 \chi_k$ --- see Fig.~\ref{fig:pairing_phases}, and Eqns.~\eqref{eq:extended_swave} through \eqref{eq:fwave}. For clarity of exposition, we will focus on the case of a single pointlike scatterer, $u_{\mu, q} = 1 / \sqrt{N}$ ($\mu = c, v$). For a spatially-extended scatterer, the QPI results we obtain below would be multiplied by an envelope function, given by the Fourier transform of the spatial profile of the scatterer.

\section{Quasiparticle interference}    \label{sec:qpi}
Assuming a spin-unpolarized tip, the signal in an STS experiment ($\dd I/\dd V$) is proportional to the local density of single-electron states summed over both spins \cite{BruusFlensberg, Capriotti2003:dwave_ripples, PeregBarnea2008:QPI, Plamadeala2010:vortices},
\begin{equation}
    n (r; \omega)
    =
    - \frac{1}{\pi} \Im 
    \left( 
    G_{\uparrow \uparrow} (r, r; \omega)
    + G_{\downarrow \downarrow} (r, r; \omega)
    \right).
\end{equation}
$G_{\sigma \sigma} (r, r'; \omega)$ is the Fourier transform with respect to time $t$ of the retarded Green function
\begin{equation}
    G_{\sigma \sigma} (r, r'; t)
    =
    -i \Theta (t)
    \langle \lbrace
    c_{r \sigma} (t), c^{\dag}_{r' \sigma} (0)
    \rbrace \rangle.
\end{equation}
For brevity, we analyze only the spin-$\uparrow$ contribution --- the spin-$\downarrow$ contribution is equivalent\footnote{The equivalence follows from time-reversal symmetry in the spin-singlet cases, and from the argument in the footnote of Sec.~\ref{subsec:clean_limit} in the triplet cases.}.

\begin{figure}[!t]
    \centering
    \includegraphics[width=0.95\columnwidth]{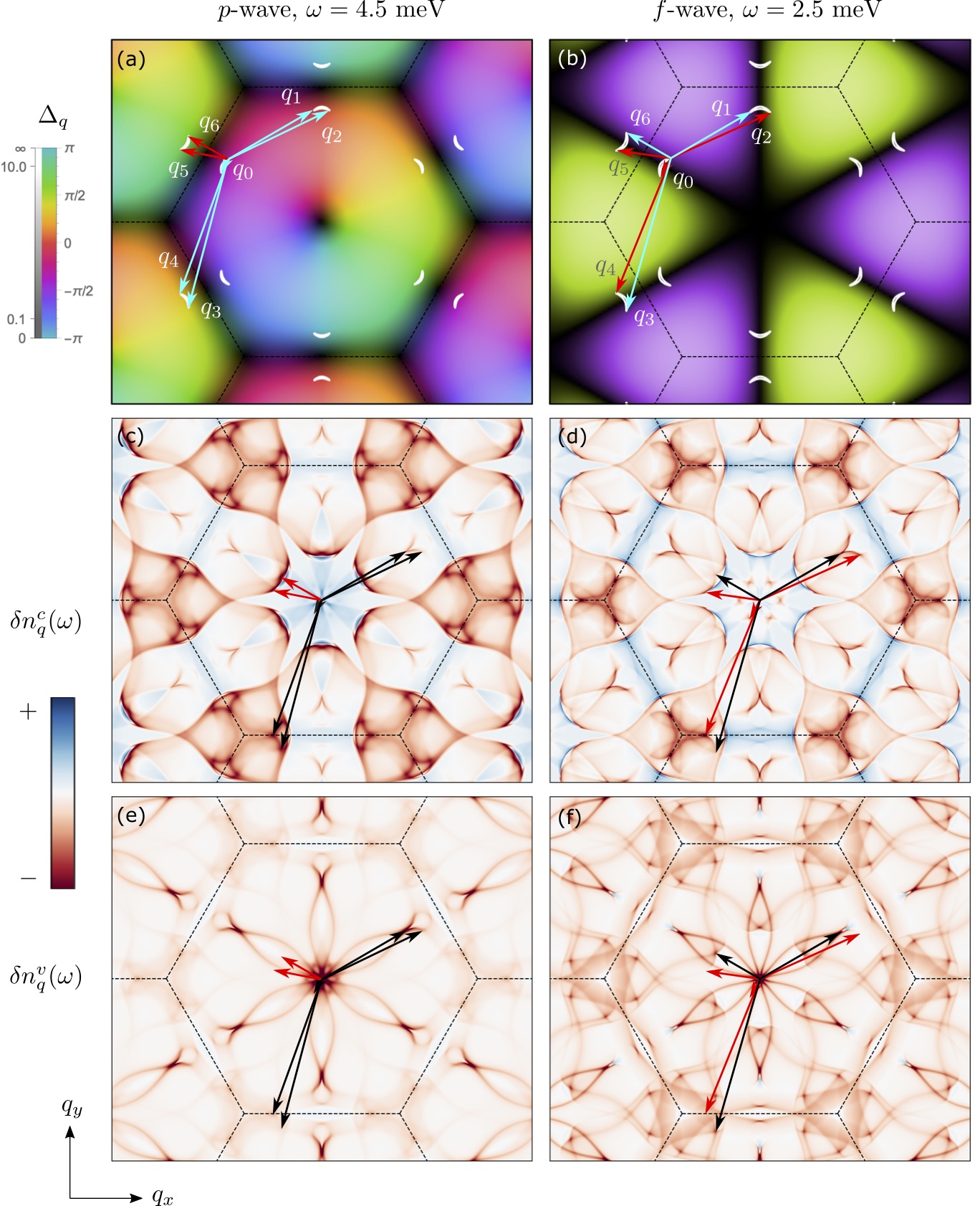}
    \caption{The spin-triplet $p$- and $f$-wave order parameters $\Delta_q$, and their QPI patterns. (a) and (b): $\Delta_q$ with amplitude and phase depicted as brightness and hue, and with the CCEs at energy $\omega$ shown as white lines. (c-f): QPI patterns for scattering from a charge impurity ((c) and (d)) and from a vortex ((e) and (f)).}
    \label{fig:pwave_vs_fwave}
\end{figure}

\begin{figure}
    \centering
    \includegraphics[width=\columnwidth]{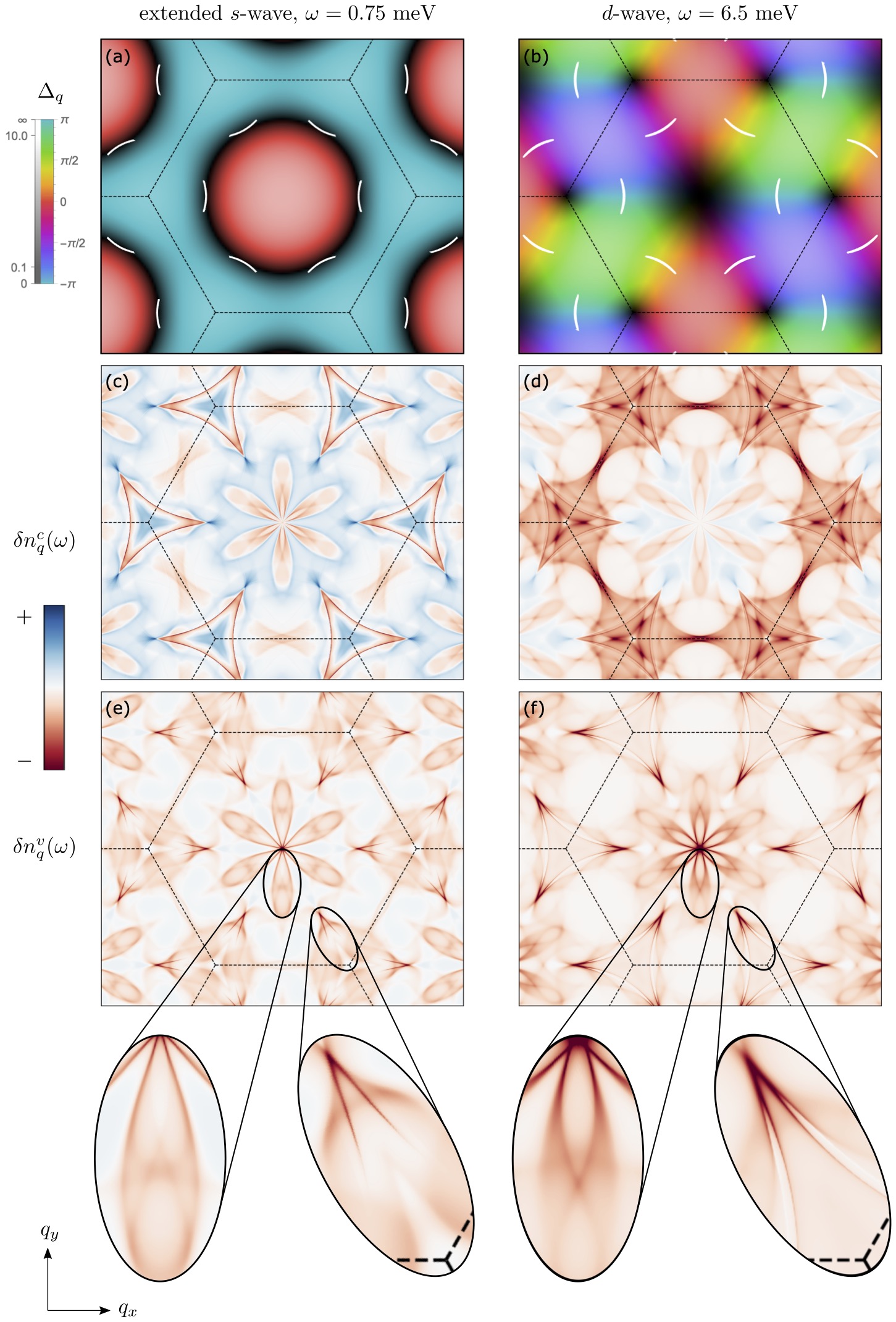}
    \caption{The spin-singlet extended-$s$- and $d$-wave order parameters $\Delta_q$, and their QPI patterns. The data are organized as in Fig.~\ref{fig:pwave_vs_fwave}.}
    \label{fig:swave_vs_dwave}
\end{figure}

Interference between scattered quasiparticle states causes spatial modulations in $n(r; \omega)$; the $q \ne 0$ part of the Fourier-transformed STS (FT-STS) signal is therefore often called the quasiparticle interference (QPI) map. As shown in Appendix \ref{appendix:QPI}, in the first Born approximation, the QPI signal for scattering from a charge impurity ($\mu = c$) or vortex ($\mu = v$) is
\begin{multline}
    \delta n^{\mu}_{\uparrow, q} [\omega]
    =
    \frac{1}{\sqrt{N}}
    \sum_{r} e^{-i q \cdot r} \delta n^{\mu}_{\uparrow} (r; \omega)  \\
    =
    - \frac{1}{\pi}
    \Im \left( \Lambda^{\mu}_{q} [\omega] \right)_{11}
\end{multline}
in terms of the $2 \times 2$ matrix
\begin{equation}    \label{eq:Lambda_def}
    \Lambda^{\mu}_{q} [\omega]
    =
    \frac{1}{N}
    \sum_k G^0_k [\omega] V^{\mu}_{k, k-q} G^0_{k-q} [\omega]
\end{equation}
where $V^{c}_{k, k-q} = \sigma_3$, $V^{v}_{k, k-q}$ is defined in Eq.~\eqref{eq:vortex_matrix}, and
\begin{equation}    \label{eq:bare_Green_function}
    G^0_k [\omega]
    =
    \left( \omega + i 0^+ - h_k \right)^{-1}
    =
    \frac{\omega + h_k}{(\omega + i0^+)^2 - E_k^2}
\end{equation}
is the clean-limit matrix-valued retarded Green function. Figs.~\ref{fig:pwave_vs_fwave} (spin-triplet order parameters) and \ref{fig:swave_vs_dwave} (spin-singlets) show $\delta n_q [\omega]$ for either type of scatterer, as well as the clean-limit contours of constant energy (CCEs) in the superconducting state. The total QPI signal is a linear combination of $\delta n^c$ and $\delta n^v$; in Fig.~\ref{fig:pwave_vs_fwave}, transitions indicated by black arrows are enhanced by increasing the applied magnetic field (which increases vortex density), while those indicated by red arrows are not. Black arrows in the bottom two rows are coloured blue in the top row for visual clarity.

Notice the clear difference between the geometry of the low-energy CCEs for the triplet order parameters ($p$- and $f$-wave, Fig.~\ref{fig:pwave_vs_fwave}) and the singlets ($s$- and $d$-wave, Fig.~\ref{fig:swave_vs_dwave}), and the correspondingly different QPI patterns. There are also subtler characteristic differences between the different angular momentum channels with the same parity, i.e.~$s$- vs.~$d$-wave and $p$- vs.~$f$-wave. We will discuss these differences in Sec.~\ref{subsec:identifying_delta}.

\subsection{QPI in the on-shell approximation}  \label{subsec:qpi_onshell}
 
The function $\Lambda^{\mu}_q [\omega]$, defined in Eq.~\eqref{eq:Lambda_def}, includes all possible single elastic scattering events at energy $\omega$ where a quasiparticle picks up momentum $q$ from the scatterer. The dominant contributions to  $\Lambda^{\mu}_q [\omega]$ come from on-shell scattering, for which $\omega^2 = E_{k}^2 = E_{k-q}^2$ (such that both propagators hit a pole) --- in other words, the dominant scattering occurs between momenta on the CCE at energy $\omega$. Writing out the summands explicitly using Eq.~\eqref{eq:bare_Green_function}, a charge impurity yields
\begin{widetext}
\begin{subequations}    \label{eq:scattering_integrands}
\begin{equation}    \label{eq:charge_integrand}
    \left( G^{0}_{k} \sigma_3 G^{0}_{k-q} \right)_{11}
    =
    \frac
    {\left( \omega + \epsilon_k \right) 
    \left( \omega + \epsilon_{k-q} \right)
    - \Delta_k \Delta_{k-q}^*}
    {\left( \left(\omega + i 0^+ \right)^2 - E_k^2 \right)
    \left( \left(\omega + i 0^+ \right)^2 - E_{k-q}^2 \right)},
\end{equation}
while a vortex yields (using $\chi = \Delta / \Delta_0$)
\begin{equation}    \label{eq:vortex_integrand}
    \left( G^{0}_{k} V^{v}_{k,k-q} G^{0}_{k-q} \right)_{11}
    =
    \frac
    {
    \frac{1}{\Delta_0} \left( \omega + \epsilon_k \right) 
    \left( \Delta_k \Delta_{k-q}^* + | \Delta_{k-q} |^2 \right) 
    +
    \frac{1}{\Delta_0} \left( | \Delta_k |^2 + \Delta_k \Delta_{k-q}^* \right)
    \left(\omega + \epsilon_{k-q} \right)
    }
    {\left( \left(\omega + i 0^+ \right)^2 - E_k^2 \right)
    \left( \left(\omega + i 0^+ \right)^2 - E_{k-q}^2 \right)}.
\end{equation}
\end{subequations}
\end{widetext}
Physically, the denominators are the same in the two cases because the on-shell condition depends only on the energy of the scattered quasiparticle, and not on the properties of the scatterer. This means that for any scatterer, strong features in the FT-STS signal are constrained to lie in the same regions of $q$-space where the on-shell condition is satisfied, i.e.~where the joint density of states for momentum transfer $q$ is large. Crucially, however, even if a scattering process is energetically on-shell, a mismatch between the states involved\footnote{Specifically, a mismatch between the particle-hole pseudospinors.} may partially or completely suppress the scattering. The numerators of Eqns.~\eqref{eq:charge_integrand} and \eqref{eq:vortex_integrand} are hence different because, roughly speaking, the intermediate states\footnote{The term ``intermediate states'' is imprecise in this context, because there are no true intermediate states in the first Born approximation. The term remains useful in the present context because the states we are probing (pure electron or pure hole states) are not the energy eigenstates of the unperturbed system, so the bare (mean-field) propagator by itself already produces transitions between them.} of the scattering process are different for the two types of scatterer. The difference originates in the fact that a charge impurity scatters particles into particles (and holes into holes), while a vortex scatters particles into holes (and vice versa).

The summand Eq.~\eqref{eq:vortex_integrand} vanishes when $\Delta_k = -\Delta_{k-q}$; the corresponding vortex scattering processes $k - q \rightarrow k$ are hence completely suppressed. In the following section, we will show how this suppression plants signatures of the order parameter symmetry into the vortex scattering QPI.

In principle, if the single-particle bandstructure were known exactly, the charge scattering signal alone would contain enough information to unambiguously identify the pairing channel --- see Figs.~\ref{fig:pwave_vs_fwave} and \ref{fig:swave_vs_dwave}. However, in reality, the single-particle bandstructure is never available with complete precision, and the QPI patterns are sensitive to details of the CCE geometry. The (out-of-plane magnetic field-dependent) vortex scattering signal provides additional information, at the cost of performing the experiment under magnetic field. We therefore propose to measure the QPI pattern as a function of magnetic field. The total signal will be a combination of charge impurity scattering and vortex scattering, $\alpha \delta n^c + \beta \delta n^v$. $\alpha$ will not have strong field dependence, but $\beta$ will increase with magnetic field, as vortices are created. As we will show in the next section, certain features in the QPI will be enhanced with the field, and others will not, depending on the symmetry of the order parameter. We note in passing that the magnetic field should be kept weak enough that vortices are sufficiently dilute to avoid forming a vortex lattice, which could introduce spurious QPI peaks associated with the periodic structure of scatterers.

\subsection{Identifying the order parameter using vortex scattering QPI}
\label{subsec:identifying_delta}

\subsubsection{The simplest nontrivial case: f-wave }
To demonstrate the utility of Eq.~\eqref{eq:vortex_integrand}, the case of $f$-wave pairing ($\ell = 3$) is especially illustrative. The $f$-wave order parameter on the triangular ($C_6$-symmetric) lattice has many features analogous to the $d$-wave order parameter on the square ($C_4$-symmetric) lattice, familiar from the cuprate superconductors \cite{PeregBarnea2008:QPI}; in both cases, the order parameter changes sign under a minimal $C_6$ (for a triangular lattice) or $C_4$ (for a square lattice) rotation, and has nodal lines. The low-energy quasiparticles then live near the intersections between the nodal lines $\Delta = 0$ and the Fermi surface $\epsilon = 0$. At small energies $\omega$, the CCEs $\omega = E = \pm \sqrt{\epsilon^2 + | \Delta |^2}$ form boomerang shapes, as shown in panel (b) of Fig.~\ref{fig:pwave_vs_fwave}.

Several features of the $f$-wave vortex scattering QPI pattern, depicted in panel (f) of Fig.~\ref{fig:pwave_vs_fwave}, follow readily. First, transitions from one ``tip'' of a boomerang to the other tip of the same boomerang (e.g.~$q_0$ in Fig.~\ref{fig:pwave_vs_fwave}) have $\Delta_k = -\Delta_{k-q}$, so those transitions are suppressed: $\Delta_k \Delta_{k-q}^* + | \Delta_{k-q} |^2 = | \Delta_k |^2 + \Delta_k \Delta_{k-q}^* = 0$. On the other hand, transitions from the tip of one boomerang to the nearest tip of a nearest-neighbour boomerang (e.g.~$q_1$) have $\Delta_k = \Delta_{k-q}$, so those transitions are not suppressed at all. Considering all six distinct boomerangs, all possible tip-to-tip transitions have $\Delta_k = \pm \Delta_{k-q}$. Those transitions for which $\Delta_{k} = -\Delta_{k-q}$ are completely suppressed, and their associated vortex QPI signals will not increase with magnetic field. Examples are indicated by red arrows in Fig.~\ref{fig:pwave_vs_fwave}. Those transitions for which $\Delta_k = \Delta_{k-q}$ are not suppressed at all, and their vortex QPI signals will increase with magnetic field. Examples are indicated by cyan (top row) or black (other rows) arrows in Fig.~\ref{fig:pwave_vs_fwave}. In contrast, the phase winding of the $p$-wave case means that several transitions which are absent from the $f$-wave vortex QPI signal are present in the $p$-wave QPI signal. We discuss this in the following section.

\subsubsection{s-, p-, and d-wave}
The $s$-, $p$-, and $d$-wave cases are slightly more complex to analyze. The extended $s$-wave order parameter has a nodal ring, which may or may not intersect the normal-state Fermi surface, depending on doping. The $p$- and $d$-wave order parameters have no nodal lines, and their phases instead wind continuously. However, in all cases, since $| \Delta |$ varies over the Fermi surface, the low-energy CCEs still form six pockets, as shown in panel (a) of Fig.~\ref{fig:pwave_vs_fwave}, and panels (a) and (b) of Fig.~\ref{fig:swave_vs_dwave}.

The singlet-versus-triplet question can be easily resolved from the QPI patterns by eye --- compare Fig.~\ref{fig:pwave_vs_fwave} to Fig.~\ref{fig:swave_vs_dwave}. We then focus on the question of distinguishing $p$- from $f$-wave, and $s$- from $d$-wave. We note that thermodynamic probes, which can detect nodes in the spectrum of Bogoliubov quasiparticles, would also be highly useful to compliment our proposed experiment.

We now turn our attention to the $p$-wave case. The order parameter and Bogoliubov CCEs are shown in panel (a) of Fig.~\ref{fig:pwave_vs_fwave}, and the QPI data are shown in panels (c) and (e). The CCEs resemble those of the $f$-wave case, taking the shape of boomerangs near the tips of the warped hexagonal Fermi surface. When the boomerangs are small (i.e.~at energies $\omega$ just above the gap), the order parameter phase $\phi$ is approximately constant over each boomerang. Since the phase winds only once as a function of angle, only those pockets which are opposite one another have a relative minus sign in the order parameter. Therefore, only the transitions at $q_5$ and $q_6$ are suppressed for the $p$-wave order parameter. Notably, signals at $q_0$ and $q_2$ will show strong magnetic field dependence for a $p$-wave order parameter, but not for $f$-wave. On the other hand, the signal at $q_6$ will show strong field-dependence for $f$-wave, but weak (if any) for $p$-wave. Our results therefore provide a sharp distinction between these two spin-triplet order parameters.

If $\omega$ is large enough that the boomerangs are no longer small, even though the approximation that $\phi$ is nearly constant over a given boomerang breaks down, a similar analysis still holds. The result is a hierarchy of transitions, which are enhanced at different rates as the magnetic field is increased. However, the signals are clearest in the small-pocket regime (where $\omega$ is chosen to probe the lowest-energy quasiparticles), so we suggest that interested experimentalists focus their attention there. The evolution of the CCEs and QPI signals with $\omega$ is shown in Appendix \ref{appendix:energy_evolution}.

In the preceding analysis, we showed how the simple sign changes in the $f$-wave order parameter result in QPI patterns with characteristic differences from the $p$-wave case. Distinguishing between singlet order parameters (e.g.~$s$- vs.~$d$-wave) using QPI alone is less clear-cut. For these order parameters, no prominent transitions have $\Delta_k = - \Delta_{k-q}$, so no such transitions are fully suppressed in the vortex QPI signal. However, the nodal ring $\Delta = 0$ in the extended $s$-wave order parameter (see Fig.~\ref{fig:swave_vs_dwave}(a)) does produce a signature in the vortex QPI, which sets the extended $s$- and $d$-wave cases apart. Transitions involving states on parts of the CCEs which fall close to the nodal ring are suppressed by the very small values of $|\Delta|$ there, suppressing some fine-scale contour-like structures in the QPI maps --- compare the zoomed regions of Fig.~\ref{fig:swave_vs_dwave}(e) and (f). 

We stress that the signatures distinguishing between the singlets are less definitive than those distinguishing between the triplets. As well, the signatures will likely be absent if on-site $s$-wave pairing is dominant (since there is no nodal ring in that case). Therefore, if the experimental QPI data suggest a singlet order parameter, other experiments should be performed to pin down its (orbital) angular momentum. $s$-wave superconductors are known to be robust against time-reversal symmetric disorder; charge impurities in such a superconductor will not bind in-gap states \cite{Balatsky}. On the other hand, higher-$\ell$ superconductors are sensitive to time-reversal symmetric disorder, and the same scanning-tunneling probes used to measure QPI could be used to observe impurity bound states at charge impurities, distinguishing the higher-$\ell$ singlets (most notably $d$-wave) from the $s$-wave case \cite{Hsiu-Hau}.

\section{Conclusion}    \label{sec:conclusion}
In this paper, we proposed using LDOS modulations as a means of identifying the order parameter symmetry in two-dimensional superconductors. We analyzed the QPI patterns which would result from a variety of pairing functions, defined on the nearest-neighbor bonds of a triangular lattice. We computed these patterns using a phenomenological tight-binding model, tailored for systems of adatoms deposited on a host surface, using the case of tin-on-silicon as a prototypical example. The QPI patterns are obtained as the Fourier-transformed LDOS modulations which would result from quasiparticles scattering off of charge impurities and superconducting vortices. The QPI signal at momentum $q$ and energy $\omega$ encodes the probability of scattering with that momentum transfer on the relevant contour of constant energy, and reveals the matrix elements of the scattering potential combined with the Green function of the clean system. We showed that certain vortex scattering events are suppressed, or even absent, depending on the particular superconducting order parameter at play. Since the vortex density is controlled by an applied out-of-plane magnetic field, the relative strength of the vortex and charge QPI signals can be tuned by such a field, allowing the two contributions to be distinguished. Therefore, our results could be used in conjunction with a future LDOS measurement in magnetic field, in order to identify the order parameter symmetry in the tin-on-silicon adatom system. Our results can also be applied to any other two-dimensional superconductor on a triangular lattice.

\section{acknowledgements}
BAL acknowledges financial support from FRQNT. TPB acknowledges financial support from NSERC and FRQNT.
\appendix

\section{Band dispersion and order parameters}
\label{appendix:dispersion+order_params}

The normal-state dispersion $\epsilon_k$ includes hopping out as far as the fifth nearest-neighbours:
\begin{equation}    \label{eq:epsilon}
\begin{aligned}
    \epsilon_k
    = &
    - 2 t_1 \Bigg[ \cos \left( k_x \right)
        + 2 \cos \left( \frac{1}{2} k_x \right)
        \cos \left( \frac{\sqrt{3}}{2} k_y \right) \Bigg]  \\
    & - 2 t_2 \Bigg[ \cos \left( \sqrt{3} k_y \right)
        + 2 \cos \left( \frac{3}{2} k_x \right)
        \cos \left( \frac{\sqrt{3}}{2} k_y \right) \Bigg]  \\
    & - 2 t_3 \Bigg[ \cos \left(2 k_x \right) 
        + 2 \cos \left(k_x \right) 
        \cos \left( \sqrt{3} k_y \right) \Bigg]    \\
    & - 4 t_4 \Bigg[ \cos \left( \frac{5}{2} k_x \right)
        \cos \left( \frac{\sqrt{3}}{2} k_y \right) \\
        & \hspace{1.5cm} + \cos \left( 2 k_x \right)
        \cos \left( \sqrt{3} k_y \right)   \\
        & \hspace{2cm} + \cos \left( \frac{1}{2} k_x \right)
        \cos \left( \frac{3 \sqrt{3}}{2} k_y \right) \Bigg]    \\
    & -2 t_5 \Bigg[ \cos \left( 2 \sqrt{3} k_y \right)
        + 2 \cos \left( 3 k_x \right) 
        \cos \left( \sqrt{3} k_y \right) \Bigg]
\end{aligned}
\end{equation}
Based on density functional theory, for the case of Sn atoms on a Si (111) surface, the authors of Ref.~\cite{Li2013:magnetic_surface_order} obtained the hopping parameters $t_1 = -52.7 \text{ meV}$, $t_2 = -0.3881 \, t_1$, $t_3 = 0.1444 \, t_1$, $t_4 = -0.0228 \, t_1$, and $t_5 = -0.0318 \, t_1$. We use these parameters in all numerics, and neglect any magnetic ordering. The dispersion $\epsilon_k$ and its Fermi surface are shown in Fig.~\ref{fig:dispersion}.

We write the superconducting order parameter as $\Delta_k = \Delta_0 \chi_k$ in terms of a dimensionless function $\chi_k$, and a parameter $\Delta_0 > 0$ which controls the size of the gap in the non-nodal cases, or the ``gap velocity" ($\sim \partial \Delta / \partial k$) in the nodal cases. We consider 
\begin{subequations}    \label{eq:order_params}
\begin{enumerate}
    \item Extended (i.e.~nearest-neighbour) $s$-wave ($\ell = 0$),
        \begin{equation}    \label{eq:extended_swave}
            \chi_{s, k}
            =
            \cos \left( k_x \right)
            + 2 \cos \left( \frac{1}{2} k_x \right)
                \cos \left( \frac{\sqrt{3}}{2} k_y \right);
        \end{equation}
    \item Chiral $p$-wave ($\ell = 1$),
        \begin{multline}    \label{eq:pwave}
            \chi_{p, k}
            =
            i \sin \left( k_x \right)
            +
            i \sin \left( \frac{1}{2} k_x \right)
            \cos \left( \frac{\sqrt{3}}{2} k_y \right)
            \\
            + \sqrt{3} \cos \left( \frac{1}{2} k_x \right)
            \sin \left( \frac{\sqrt{3}}{2} k_y \right);
        \end{multline}
    \item Chiral $d$-wave ($\ell = 2$),
        \begin{multline}    \label{eq:dwave}
            \chi_{d, k}
            =
            \cos \left( k_x \right)
            - \cos \left( \frac{1}{2} k_x \right)
            \cos \left( \frac{\sqrt{3}}{2} k_y \right)
            \\
            + i \sqrt{3} 
            \sin \left( \frac{1}{2} k_x \right)
            \sin \left( \frac{\sqrt{3}}{2} k_y \right);
        \end{multline}
    \item $f$-wave ($\ell = 3$),
        \begin{multline}    \label{eq:fwave}
            \chi_{f, k}
            =
            i \sin \left( k_x \right)
            - 2 i \sin \left( \frac{1}{2} k_x \right)
            \cos \left( \frac{\sqrt{3}}{2} k_y \right).
        \end{multline}
\end{enumerate}
\end{subequations}
In numerics, we use the somewhat large value $\Delta_0 = 5$ meV, to slightly exaggerate features in the CCEs for clarity of presentation.

\section{Quasiparticle interference in the first Born approximation}
\label{appendix:QPI}
STS measures the local density of electronic states, which is given (for a spinless model) by \cite{BruusFlensberg,Capriotti2003:dwave_ripples}
\begin{equation}
    n (r; \omega)
    =
    - \frac{1}{\pi} \Im G (r, r; \omega).
\end{equation}
Taking the Fourier transform,
\begin{equation}
    n_{q}
    =
    \frac{1}{\sqrt{N}} \sum_{r}
    e^{- i q \cdot r} n (r).
\end{equation}
Now express $G(r,r)$ in terms of its Fourier transform with respect to both indices:
\begin{equation}
    G(r, r)
    =
    \frac{1}{N} \sum_{k k'} e^{i (k - k') \cdot r} G_{k k'},
\end{equation}
so
\begin{multline}    \label{eq:general_n_q}
    n_{q}
    =
    -\frac{1}{\pi}
    \frac{1}{\sqrt{N^3}} \sum_{r} \sum_{k k'}
    e^{- i q \cdot r}
    \Im \left( e^{i (k - k') \cdot r} G_{k k'} \right)  \\
    =
    -\frac{1}{\pi}
    \frac{1}{\sqrt{N^3}} \sum_{r} \sum_{k k'}
    e^{- i q \cdot r}
    \big\lbrace
    \cos \left( (k - k') \cdot r \right) \Im G_{k k'}   \\
    + i \sin \left( (k - k') \cdot r \right) \Re G_{k k'}
    \big\rbrace \\
    =
    -\frac{1}{2 \pi}
    \frac{1}{\sqrt{N}} \sum_{k} 
    \big\lbrace
    \Im \left( G_{k, k-q} + G_{k, k+q} \right) \\
    + \Re \left( G_{k, k-q} - G_{k, k+q} \right)
    \big\rbrace.
\end{multline}

To proceed any further, we need to compute the Green function $G_{k k'}$ including scattering. The simplest approach is the first Born approximation, and the calculation is easiest using the Nambu formalism, in terms of the two-component particle-hole pseudospinors given by Eqns.~\eqref{eq:singlet_spinor} and \eqref{eq:triplet_spinor}. We consider a single pointlike scatterer. For a charge impurity ($\mu = c$) or a vortex ($\mu = v$), we have
\begin{equation}
    G_{k k'}
    =
    G^0_{k} \delta_{k k'}
    + \frac{1}{\sqrt{N}} G^0_{k} V^{\mu}_{k k'} G^0_{k'}.
\end{equation}
Recall that $V^{c}_{k k'} = \sigma_3$ and $V^{v}_{k k'} = \begin{pmatrix} 0 & \chi_{k} + \chi_{k'} \\ \chi_{k}^* + \chi_{k'}^* & 0 \end{pmatrix}$. Note that the dispersion has inversion symmetry in the plane, i.e.~$\epsilon_k = \epsilon_{-k}$. For the spin-singlet pairing channels, $\chi_{k} = \chi_{-k}$ as well, so $h_{k} = h_{-k}$, $G^0_{k} = G^0_{-k}$, and $V^{v}_{k, k'} = V^{v}_{-k, k'} = V^{v}_{k, -k'}$. Altogether, we can then simplify
\begin{multline}
    \sum_{k} G^0_{k} V^{\mu}_{k, k+q} G^0_{k+q}
    =
    \sum_{k} G^0_{-k} V^{\mu}_{-k, -k-q} G^0_{-k-q}   \\
    =
    \sum_{k} G^0_{k} V^{\mu}_{k, k-q} G^0_{k-q}
\end{multline}
where we relabelled $k \rightarrow -k$ to obtain the final equality. Eq.~\eqref{eq:general_n_q} then simplifies to
\begin{equation}    \label{eq:n_lambda}
    \delta n^{\mu}_{\uparrow q}
    =
    - \frac{1}{\pi} \Im \Lambda^{\mu}_{q}
\end{equation}
with
\begin{equation}
    \Lambda^{\mu}_{q}
    =
    \frac{1}{N}
    \sum_{k} G^0_{k} V^{\mu}_{k, k-q} G^0_{k-q}.
\end{equation}
For the spin-triplet cases, the order parameter is antisymmetric under in-plane inversion: $\chi_k = - \chi_{-k}$, which means that inversion symmetry acts on $h$, $V^{v}$, and $G^0$ as $h_{k} = \sigma_3 h_{-k} \sigma_3$, $V^{v}_{k, k'} = \sigma_3 V^{v}_{-k, -k'} \sigma_3 = -V^{v}_{-k, -k'}$, and $G^0_{k} = \sigma_3 G^0_{-k} \sigma_3$. Inversion acts trivially on $V^{c} = \sigma_3$, but we can still write it as $\sigma_3 = \sigma_3 \sigma_3 \sigma_3$. Then, we can simplify 
\begin{multline}
    \sum_{k} G^0_{k} V^{\mu}_{k, k+q} G^0_{k+q}
    =
    \sum_{k} \sigma_3 G^0_{-k} V^{\mu}_{-k, -k-q} G^0_{-k-q} \sigma_3  \\
    =
    \sum_{k} \sigma_3 G^0_{k} V^{\mu}_{k, k-q} G^0_{k-q} \sigma_3,
\end{multline}
which yields
\begin{multline}
    \delta n^{\mu}_{\uparrow q}
    =
    - \frac{1}{2 \pi} 
    \Big\lbrace
    \Im \left( \Lambda^{\mu}_{q}
    + \Lambda^{\mu}_{-q} \right)_{11}
    + \Re \left( \Lambda^{\mu}_{q}
    - \Lambda^{\mu}_{-q} \right)_{11}
    \Big\rbrace \\
    =
    - \frac{1}{2 \pi} 
    \Big\lbrace
    \Im \left( \Lambda^{\mu}_{q}
    + \sigma_3 \Lambda^{\mu}_{q} \sigma_3 \right)_{11}
    + \Re \left( \Lambda^{\mu}_{q}
    - \sigma_3 \Lambda^{\mu}_{q} \sigma_3 \right)_{11}
    \Big\rbrace \\
    =
    - \frac{1}{\pi} 
    \Im \left( \Lambda^{\mu}_{q} \right)_{11},
\end{multline}
just as in the spin-singlet cases. It is straightforward to repeat this calculation with a spatially-extended scatterer, described by a real-valued symmetric envelope function $u_{\mu} (r) = u_{\mu} (-r)$ (with Fourier transform $u_{\mu, q} = u_{\mu, -q} \in \mathbb{R}$), yielding the general result \cite{TamiPseudogapQPI}
\begin{equation}
    \delta n_{\uparrow, q}^{\mu}
    =
    - \frac{1}{\pi} u_q
    \Im \left( \Lambda^{\mu}_{q} \right)_{11}.
\end{equation}

\section{Energy-evolution of the QPI patterns}
\label{appendix:energy_evolution}

As the CCEs evolve with the probing energy $\omega$, so too do the QPI patterns. In this appendix we show the evolution through the low-energy regime.

\begin{figure*}
    \centering
    \includegraphics[width=0.8\textwidth]{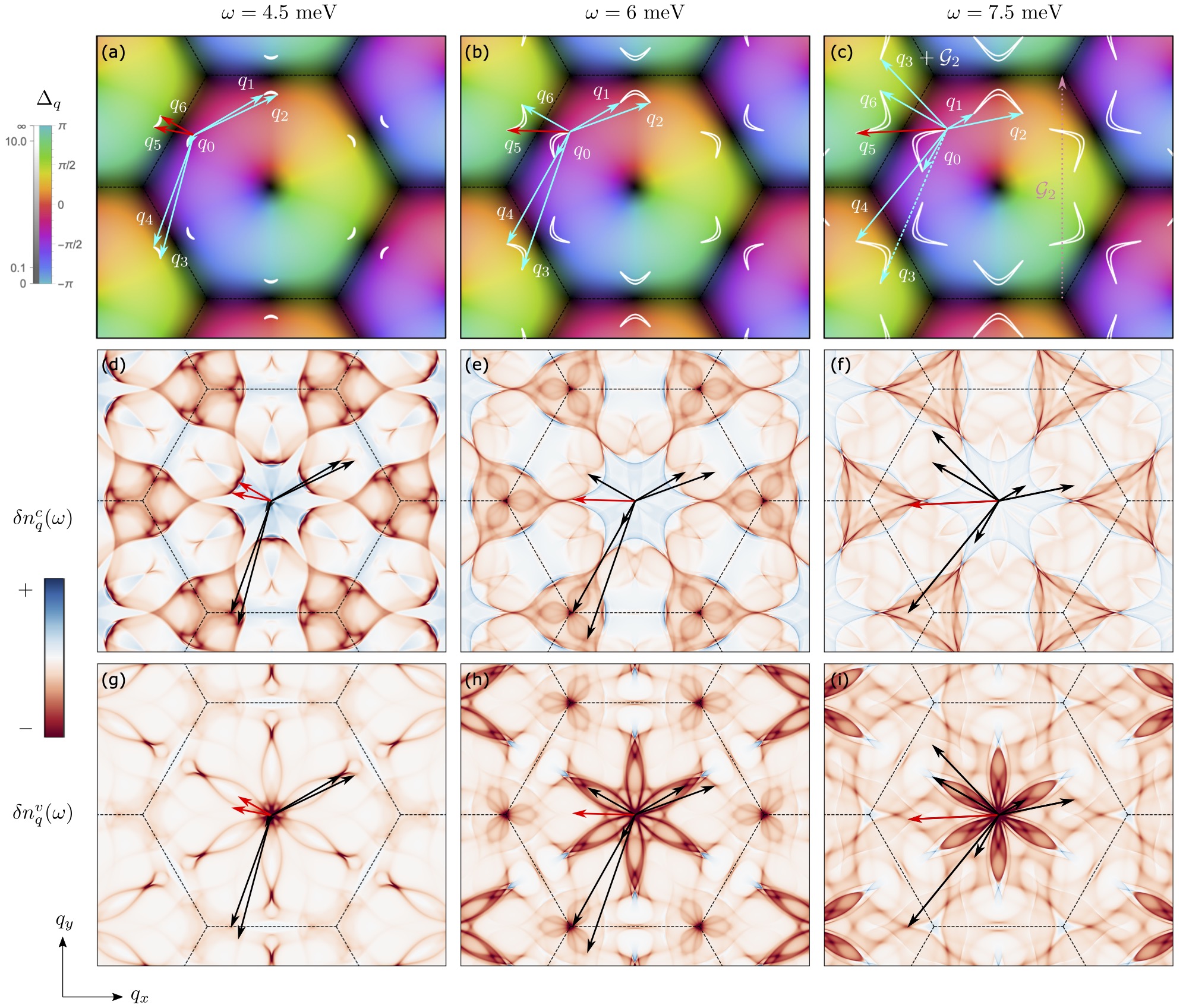}
    \caption{\label{fig:pwave_qpi_energyevolution} Energy evolution of the $p$-wave QPI.}
\end{figure*}

\begin{figure*}
    \centering
    \includegraphics[width=0.8\textwidth]{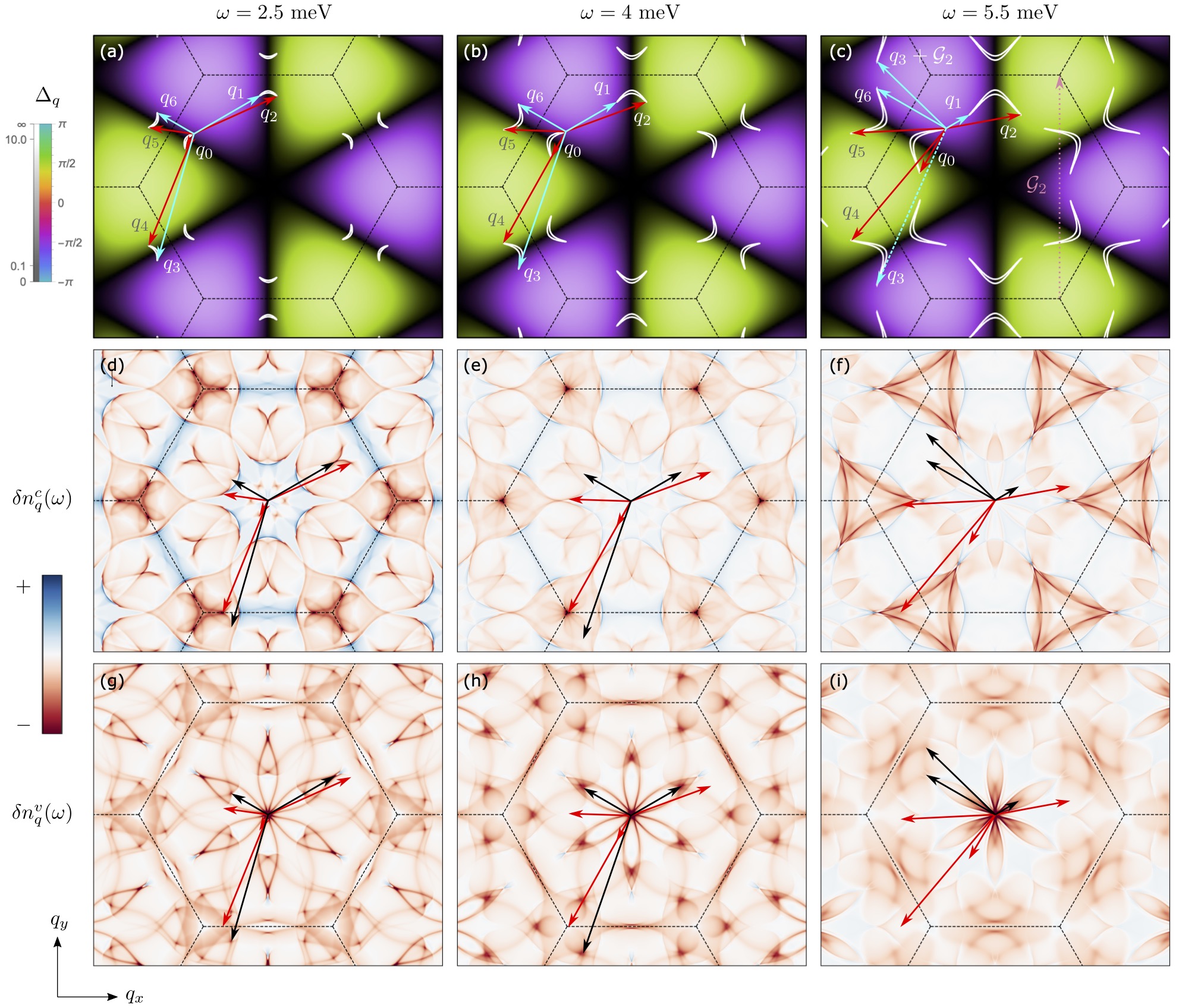}
    \caption{\label{fig:fwave_qpi_energyevolution} Energy evolution of the $f$-wave QPI.}
\end{figure*}

\begin{figure*}
    \centering
    \includegraphics[width=0.8\textwidth]{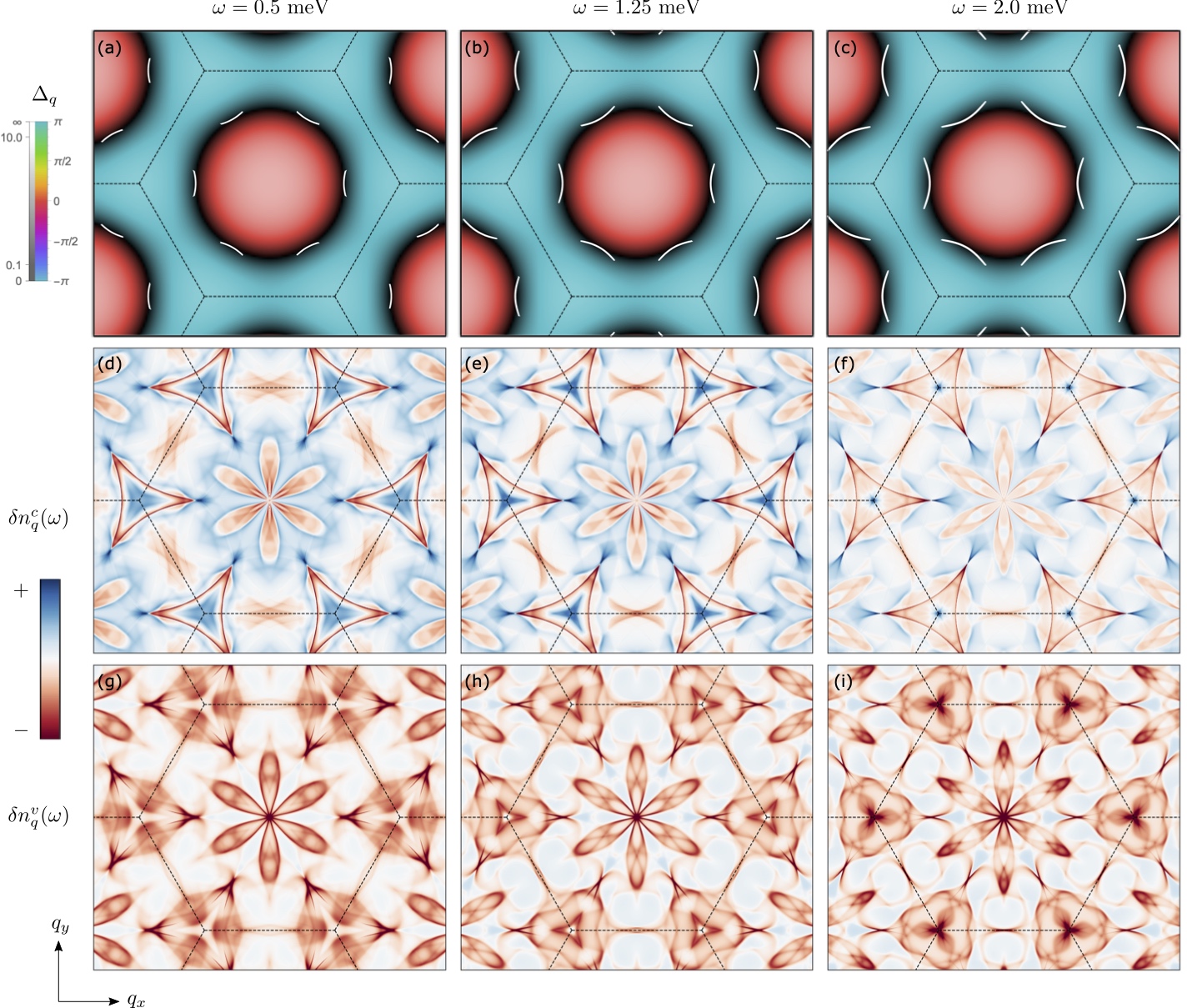}
    \caption{\label{fig:swave_qpi_energyevolution} Energy evolution of the $s$-wave QPI.}
\end{figure*}

\begin{figure*}
    \centering
    \includegraphics[width=0.8\textwidth]{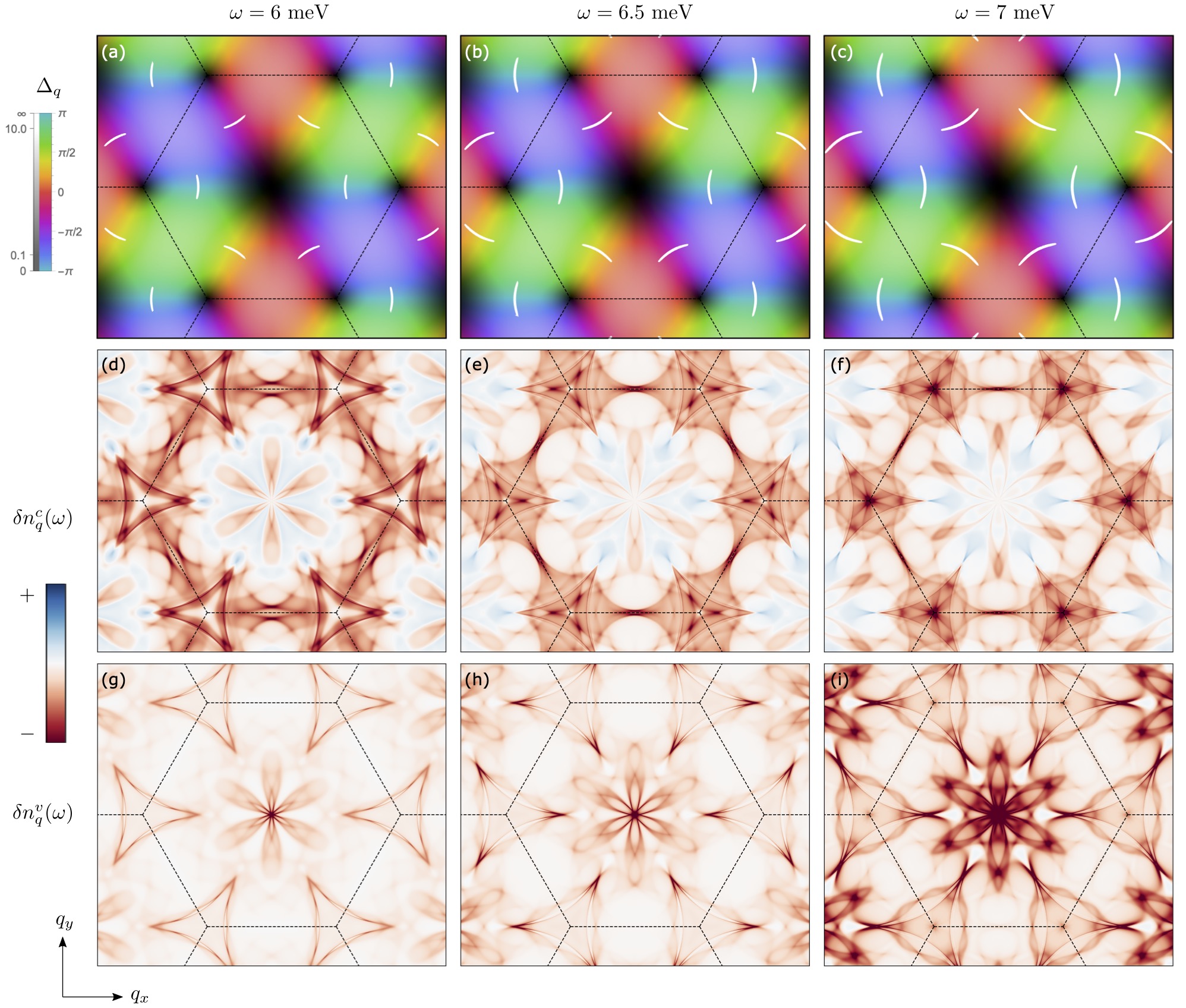}
    \caption{\label{fig:dwave_qpi_energyevolution} Energy evolution of the $d$-wave QPI.}
\end{figure*}

\bibliographystyle{apsrev4-2}
\bibliography{qpi}

\end{document}